\documentstyle[preprint,eqsecnum,aps]{revtex}
\begin{document}
\draft
\preprint{}
\title{Post-Newtonian Approximation for Spinning Particles}
\author{
H. T. Cho \footnote[1]{htcho@mail.tku.edu.tw }}
\address{Tamkang University, Department of Physics,\\
Tamsui, Taipei, TAIWAN R.O.C.}
\maketitle
\begin{abstract}
Using an energy-momentum tensor for spinning particles
due to Dixon and Bailey-Israel, we develop the 
post-Newtonian approximation for N spinning particles
in a self-contained manner. The equations of motion are 
derived directly from this energy-momentum tensor.
Following the formalism of Epstein and Wagoner, we
also obtain the waveform and the luminosity of the gravitational
wave generated by these particles.
\end{abstract}
\pacs{}

\section{Introduction}

The possibility of detecting gravitational waves from coalescences
of compact binaries in the near future by LIGO \cite{AAD} and 
VIRGO \cite{BFS} laser interferometric detectors has aroused quite 
a lot of research in this area. One of the main tools in analyzing
these situations and in obtaining the gravitational waveforms generated
is the post-Newtonian approximation \cite{CMW}. As one gets closer to 
the coalescence, higher order terms in this approximation have to be
taken into account to get results accurate enough for comparison with
future observational data.

When higher order terms are considered, spin and higher multipole
effects may be of significance, especially for neutron stars and 
Kerr black holes. For these compact objects, it is estimated that
the spin-orbit and spin-spin effects are of (post)$^{3/2}$-Newtonian
and (post)$^{2}$-Newtonian order, respectively \cite{KWW}. Therefore,
to get accurate results to these orders, spin effects must be included.

A number of authors have studied these spin effects in 
\cite{KWW}-$\!\!$\cite{LEK}.
In \cite{LEK}, both the spin-orbit and spin-spin effects
are considered in details using the multipole formalism of Blanchet,
Damour, and Iyer (BDI) \cite{BD,DI}. The BDI formalism is rigorous, but
rather complicated in which each body is regarded as a spherically
symmetric, rigidly rotating perfect fluid.

In this paper we would like to introduce another approach which is 
simpler and self-contained in the sense that the equations of motion
are also derived along the way. In this approach we still regard the
bodies as point particles, but with structures such as spin. The central
quantity here is the energy-momentum tensor for these particles. In
general spacetimes this energy-momentum tensor was first introduced by
Dixon \cite{WGD}, and was 
later elaborated on by Bailey and Israel (BI) \cite{BI}. 
In \cite{BI}, the Lagrangian density of a particle is expanded in a 
covariant Taylor expansion about another world-line, presumably the 
center-of-mass of some extended object. Higher order terms in this
expansion correspond to spin, tidal effects and so on. From this
Lagrangian one can obtain the energy-momentum tensor in the usual fashion.
Therefore, in this formalism one can include structures such as spin,
quadrupole moment and so on in a systematic manner. In this paper,
however, we shall concentrate only on the spin effects.

With this energy-momentum tensor for particles with spin, one can derive
the equations of motion for spinning particles, that is, the Papapetrou
equations, by requiring this energy-momentum tensor to be symmetric and
conserving. This will be done in the next section. In Section III, we
develop the post-Newtonian approximation for N spinning particles using
this tensor. We follow closely the procedures in \cite{SW}. 
As an illustration of how our approach works, 
we choose the simpler case that the spin-orbit
and spin-spin effects are both of post-Newtonian order. In this case, 
we are not dealing with compact objects, but ordinary ones. The equations
of motion for positions and spins are expanded up to (post)$^{2}$-Newtonian
order. In Section IV we use the formalism of Epstein and Wagoner (EW)
\cite{EW} to obtain the waveform and the luminosity 
of the gravitational wave generated by the motion of these N 
spinning particles up to post-Newtonian order.
This can be considered as an extension of the result of Wagoner and Will
\cite{WW} to spinning particles. Conclusions and discussions are given in
Section V.

\section{Energy-momentum tensor}

In \cite{BI}, BI devised a general method to derive
the energy-momentum tensor $T^{\mu\nu}$
for particles with structures such as spin.
First the Lagrangian for a charged point particle is expanded covariantly
about another world-line. For the particle to represent part of an 
extended object, this world-line is usually chosen to be the center-of-mass
of the object. However, their formalism is rather general that one can 
actually choose any convenient world-line. The higher order terms in this 
expansion can be identified with the structures of the particle
corresponding to higher gravitational and electromagnetic multipole
moments. With this Lagrangian expansion, one can obtain the equations
of motion and $T^{\mu\nu}$ by the usual variations.

Here we concentrate on particles with the extra structure of spin, which
is related to the gravitational dipole moment. In this case, $T^{\mu\nu}$
for a spinning particle is given by \cite{BI},
\begin{equation}
T^{\mu\nu}=T^{\mu\nu}_{(can)}+\nabla_{\rho}B^{\mu\nu\rho},
\end{equation}
where
\begin{eqnarray}
T^{\mu\nu}_{(can)}&=&\frac{1}{\sqrt{-g}}\int d\tau p^{\mu}v^{\nu}
\delta^{4}(x-z(\tau)),
\\
B^{\mu\nu\rho}&=&-\frac{1}{2}(S^{\mu\nu\rho}+S^{\rho\mu\nu}
+S^{\rho\nu\mu}),
\\
S^{\mu\nu\rho}&=&\frac{1}{\sqrt{-g}}\int d\tau S^{\mu\nu}
v^{\rho}\delta^{4}(x-z(\tau)),
\end{eqnarray}
and $p^{\mu}$ is the canonical momentum, $S^{\mu\nu}$ the spin tensor.
$T^{\mu\nu}_{(can)}$ is therefore the canonical energy-momentum tensor,
and $B^{\mu\nu\rho}$ is the Belinfante tensor due to spin. The equations
of motion can be derived in the following manner. First we require 
$T^{\mu\nu}$ to be symmetric in $\mu$ and $\nu$, 
$T^{\mu\nu}=T^{\nu\mu}$, which gives
\begin{equation}
\frac{1}{\sqrt{-g}}\int d\tau\ (p^{\mu}v^{\nu}
-p^{\nu}v^{\mu})\delta^{4}(x-z(\tau))
-\nabla_{\rho}\left[\frac{1}{\sqrt{-g}}\int d\tau S^{\mu\nu}
v^{\rho}\delta^{4}(x-z(\tau))\right]=0.
\label{sym}
\end{equation}
Using the formulae,
\begin{equation}
\partial_{\mu}\sqrt{-g}=\sqrt{-g}\ \Gamma^{\rho}_{\rho\mu},
\label{form1}
\end{equation}
and
\begin{equation}
v^{\mu}\partial_{\mu}\delta^{4}(x-z(\tau))
=-\frac{\partial}{\partial\tau}\delta^{4}(x-z(\tau)),
\label{form2}
\end{equation}
the derivative becomes
\begin{equation}
\nabla_{\rho}\left[\frac{1}{\sqrt{-g}}\int d\tau S^{\mu\nu}v^{\rho}
\delta^{4}(x-z(\tau))\right]=
\frac{1}{\sqrt{-g}}\int d\tau\frac{DS^{\mu\nu}}{D\tau}
\delta^{4}(x-z(\tau)),
\end{equation}
where we have ignored surface terms. Here
\begin{eqnarray}
\frac{DS^{\mu\nu}}{D\tau}&\equiv&v^{\alpha}\nabla_{\alpha}S^{\mu\nu}
\nonumber\\
&=&\frac{dS^{\mu\nu}}{d\tau}+v^{\alpha}\Gamma^{\mu}_{\alpha\rho}
S^{\rho\nu}+v^{\alpha}\Gamma^{\nu}_{\alpha\rho}S^{\mu\rho}.
\end{eqnarray}
Therefore, Eq.(\ref{sym}) becomes
\begin{eqnarray}
&&\frac{1}{\sqrt{-g}}\int d\tau\left[\frac{DS^{\mu\nu}}{D\tau}
-(p^{\mu}v^{\nu}-p^{\nu}v^{\mu})\right]\delta^{4}(x-z(\tau))=0
\nonumber\\
&\Rightarrow&\frac{DS^{\mu\nu}}{D\tau}
=p^{\mu}v^{\nu}-p^{\nu}v^{\mu}.
\label{papa1}
\end{eqnarray}
This is the first set of the Papapetrou equations, 
the equations of motion for spinning particles.

Next we require that $T^{\mu\nu}$ is conserving,
$\nabla_{\mu}T^{\mu\nu}=0$,
which gives
\begin{equation}
\nabla_{\mu}T^{\nu\mu}_{(can)}+
\nabla_{\mu}\nabla_{\rho}B^{\nu\mu\rho}=0,
\label{con}
\end{equation}
where we have used the symmetry property of 
$T^{\mu\nu}$. Using the formulae Eqs.(\ref{form1}) and 
(\ref{form2}) again, 
we have
\begin{eqnarray}
\nabla_{\mu}T^{\nu\mu}_{(can)}&=&
\nabla_{\mu}\left[\frac{1}{\sqrt{-g}}\int d\tau p^{\nu}v^{\mu}
\delta^{4}(x-z(\tau))\right]
\nonumber\\
&=&\frac{1}{\sqrt{-g}}\int d\tau\left(\frac{Dp^{\nu}}{D\tau}\right)
\delta^{4}(x-z(\tau)).
\end{eqnarray}
For $\nabla_{\mu}\nabla_{\rho}B^{\nu\mu\rho}$, we use the identity
\begin{equation}
(\nabla_{\mu}\nabla_{\nu}-\nabla_{\nu}\nabla_{\mu})
\phi^{\alpha\beta\cdots}={R^{\alpha}}_{\xi\mu\nu}\phi^{\xi\beta\cdots}
+{R^{\beta}}_{\xi\mu\nu}\phi^{\alpha\xi\cdots}
\end{equation}
for the tensor $\phi^{\alpha\beta\cdots}$ and the symmetry properties
of the Riemann tensor $R_{\mu\nu\alpha\beta}$ to get
\begin{equation}
\nabla_{\mu}\nabla_{\rho}B^{\nu\mu\rho}
=\frac{1}{2}{R^{\nu}}_{\mu\rho\sigma}S^{\rho\sigma\mu}.
\end{equation}
Finally, Eq.(\ref{con}) becomes
\begin{eqnarray}
&&\frac{1}{\sqrt{-g}}\int d\tau\left[
\left(\frac{Dp^{\nu}}{D\tau}\right)
+\frac{1}{2}
{R^{\nu}}_{\mu\rho\sigma}S^{\rho\sigma}v^{\mu}\right]
\delta^{4}(x-z(\tau))=0
\nonumber\\
&\Rightarrow&\frac{Dp^{\nu}}{D\tau}=-\frac{1}{2}
{R^{\nu}}_{\mu\rho\sigma}S^{\rho\sigma}v^{\mu}.
\label{papa2}
\end{eqnarray}
This is the second set of the Papapetrou equations. If the spin tensor 
vanishes,
\begin{equation}
\frac{Dp^{\nu}}{D\tau}=0,
\end{equation}
which is just the geodesic equation for a point particle. Also from
Eq.(\ref{papa1}), we see that the canonical momentum can be written as 
\begin{equation}
p^{\mu}=mv^{\mu}-v_{\nu}\frac{DS^{\mu\nu}}{D\tau}.
\end{equation}

The spin tensor $S^{\mu\nu}$ is anti-symmetric in $\mu$ and $\nu$,
giving six independent components. Three of them can be eliminated by
the ``spin supplementary condition",
\begin{equation}
S^{\mu\nu}v_{\nu}=0.
\end{equation}
For $\mu=i$,
\begin{eqnarray}
&&S^{i0}v_{0}+S^{ij}v_{j}=0
\nonumber\\
&\Rightarrow &
S^{0i}=\left(\frac{v_{j}}{v_{0}}\right)S^{ij}.
\end{eqnarray}
The $\mu=0$ equation is then satisfied automatically. The remaining
three independent components of $S^{ij}$ can be written as a vector,
\begin{equation}
S^{i}\equiv\frac{1}{2}\epsilon^{ijk}S^{jk},
\end{equation}
which is the spin vector of the particle.

\section{Post-Newtonian approximation}

In this section we develop the post-Newtonian expansion for N spinning
particles using the energy-momentum tensor $T^{\mu\nu}$
introduced in the last 
section. We follow closely the procedure in \cite{SW}. In the post-Newtonian
approximation, it is assumed that the velocities of the particles are 
small,
\begin{equation}
v\sim\epsilon^{1/2}\ll 1.
\end{equation}
Moreover, the gravitational potential $\phi$ is comparable to the kinetic
energy,
\begin{equation}
\phi\sim v^{2}\sim\epsilon.
\end{equation}
Here we take the spin $S^{i}$ to be of order $\epsilon^{1/2}$, the same
as velocity. Under this assumption, the spin-orbit and spin-spin effects
are both of order $\epsilon$. This is valid for ordinary
situations but not compact objects for which these effects will come in
only at higher orders. In this paper we wish to treat the 
former case, which is simpler, as an illustration of how our approach
works. 

To begin we expand the spacetime metric,
\begin{eqnarray}
g_{00}&=&-1+g_{00}^{(2)}+g_{00}^{(4)}+\cdots,
\\
g_{0i}&=&g_{0i}^{(3)}+\cdots,
\\
g_{ij}&=&\delta_{ij}+g_{ij}^{(2)}+\cdots,
\end{eqnarray}
where
\begin{eqnarray}
g_{00}^{(2)}&=&-2\phi,
\\
g_{00}^{(4)}&=&-2(\phi^{2}+\psi),
\\
g_{0i}^{(3)}&=&\zeta_{i},
\\
g_{ij}^{(2)}&=&-2\phi\delta_{ij}.
\end{eqnarray}
Here the superscript denotes the order in the post-Newtonian expansion,
for example, $g^{(i)}_{00}$ is the (post)$^{i/2}$-Newtonian order of 
$g_{00}$. Through the Einstein equations the potentials $\phi$, 
$\zeta_{i}$, and $\psi$ can be expressed in terms of the various 
components of $T^{\mu\nu}$ in the appropriate order,
\begin{eqnarray}
\phi(\vec{x},t)&=&-\int d^{3}\! x'\frac{(T^{00}(\vec{x}',t))^{(0)}}
{\vert\vec{x}-\vec{x}'\vert},
\label{phi}
\\
\zeta_{i}(\vec{x},t)&=&-4\int d^{3}\! x'\frac{(T^{0i}(\vec{x}',t))^{(1)}}
{\vert\vec{x}-\vec{x}'\vert},
\label{zeta}
\\
\psi(\vec{x},t)&=&\frac{\partial^{2}\chi}{\partial t^{2}}-
\int d^{3}\! x'\left[(T^{00}(\vec{x}',t))^{(2)}
+(T^{ii}(\vec{x}',t))^{(2)}\right],
\label{psi}
\end{eqnarray}
where $\nabla^{2}\chi=\phi$. Note that we have used the harmonic 
gauge \cite{SW},
\begin{equation}
g^{\mu\nu}\Gamma^{\lambda}_{\mu\nu}=0.
\end{equation}
in deriving the above expressions.

As for $T^{\mu\nu}$, we use the one given in the last
section for N spinning particles,
\begin{equation}
T^{\mu\nu}=T^{\mu\nu}_{(can)}+\nabla_{\rho}B^{\mu\nu\rho},
\end{equation}
where
\begin{eqnarray}
T^{\mu\nu}_{(can)}&=&\frac{1}{\sqrt{-g}}\sum_{A}
p^{\mu}_{A}v^{\nu}_{A}\delta^{3}(\vec{x}-\vec{x}_{A}),
\\
B^{\mu\nu\rho}&=&-\frac{1}{2}(S^{\mu\nu\rho}+
S^{\rho\mu\nu}+S^{\rho\nu\mu}),
\\
S^{\mu\nu\rho}&=&\frac{1}{\sqrt{-g}}\sum_{A}
S^{\mu\nu}_{A}v^{\rho}_{A}\delta^{3}(\vec{x}-\vec{x}_{A}),
\\
p^{\mu}_{A}&=&m_{A}v^{\mu}_{A}-v_{A\nu}
\frac{DS^{\mu\nu}_{A}}{D\tau_{A}},
\end{eqnarray}
and $A=1,...,N$. Assuming that the spin part is of the order of
$\epsilon^{1/2}$, we can expand $T^{00}$, $T^{0i}$, $T^{i0}$ and
$T^{ij}$,
\begin{eqnarray}
(T^{00})^{(0)}&=&\sum_{A}m_{A}\delta^{3}(\vec{x}-\vec{x}_{A}),
\label{t000}
\\
(T^{00})^{(2)}&=&\sum_{A}m_{A}(\phi_{A}+\frac{1}{2}v^{2}_{A})
\delta^{3}(\vec{x}-\vec{x}_{A})
-\sum_{A}\epsilon^{ijk}v^{j}_{A}S^{k}_{A}\partial_{i}
\delta^{3}(\vec{x}-\vec{x}_{A}),
\\
(T^{0i})^{(1)}&=&\sum_{A}m_{A}v^{i}_{A}\delta^{3}(\vec{x}-\vec{x}_{A})
+\frac{1}{2}\sum_{A}\epsilon^{ijk}S^{k}_{A}\partial_{j}
\delta^{3}(\vec{x}-\vec{x}_{A})
\nonumber\\
&=&(T^{i0})^{(1)},
\label{t0i1}
\\
(T^{0i})^{(3)}&=&\sum_{A}m_{A}(\phi_{A}+\frac{1}{2}v^{2}_{A})
v^{i}_{A}\delta^{3}(\vec{x}-\vec{x}_{A})
+\sum_{A}\epsilon^{ijk}v^{j}_{A}
\left(\frac{dS^{k}_{A}}{dt}\right)^{(2)}
\delta^{3}(\vec{x}-\vec{x}_{A})
\nonumber\\
&&\ +\sum_{A}\epsilon^{ijk}
\left(\frac{dv^{j}_{A}}{dt}\right)^{(2)}S^{k}_{A}
\delta^{3}(\vec{x}-\vec{x}_{A})
+\sum_{A}\epsilon^{ijk}S^{k}_{A}\phi_{A}\partial_{j}
\delta^{3}(\vec{x}-\vec{x}_{A})
\nonumber\\
&&\ -\frac{1}{2}\sum_{A}\epsilon^{ikl}v^{j}_{A}v^{k}_{A}S^{l}_{A}
\partial_{j}\delta^{3}(\vec{x}-\vec{x}_{A})
-\frac{1}{2}\sum_{A}\epsilon^{jkl}v^{i}_{A}v^{k}_{A}S^{l}_{A}
\partial_{j}\delta^{3}(\vec{x}-\vec{x}_{A})
\nonumber\\
&=&(T^{i0})^{(3)}+\sum_{A}\epsilon^{ijk}v^{j}_{A}
\left(\frac{dS^{k}_{A}}{dt}\right)^{(2)}
\delta^{3}(\vec{x}-\vec{x}_{A}),
\label{t0i3}
\\
(T^{ij})^{(2)}&=&\sum_{A}m_{A}v^{i}_{A}v^{j}_{A}
\delta^{3}(\vec{x}-\vec{x}_{A})
-\frac{1}{2}\sum_{A}\epsilon^{ijk}
\left(\frac{dS^{k}_{A}}{dt}\right)^{(2)}
\delta^{3}(\vec{x}-\vec{x}_{A})
\nonumber\\
&&\ +\sum_{A}\epsilon^{(ikl}v^{j)}_{A}S^{l}_{A}\partial_{k}
\delta^{3}(\vec{x}-\vec{x}_{A}),
\\
(T^{ij})^{(4)}&=&\sum_{A}m_{A}(\phi_{A}+\frac{1}{2}v^{2}_{A})
v^{i}_{A}v^{j}_{A}\delta^{3}(\vec{x}-\vec{x}_{A})
-\frac{1}{2}\sum_{A}\epsilon^{ijk}
\left(\frac{dS^{k}_{A}}{dt}\right)^{(4)}
\delta^{3}(\vec{x}-\vec{x}_{A})
\nonumber\\
&&\ +\sum_{A}\epsilon^{(ikl}v^{k}_{A}S^{l}_{A}
\left(\frac{dv^{j)}_{A}}{dt}\right)^{(2)}
\delta^{3}(\vec{x}-\vec{x}_{A})
+\sum_{A}\epsilon^{(ikl}
\left(\frac{dv^{k}_{A}}{dt}\right)^{(2)}S^{l}_{A}v^{j)}_{A}
\delta^{3}(\vec{x}-\vec{x}_{A})
\nonumber\\
&&\ +\sum_{A}\epsilon^{ikl}
\left(\frac{dv^{k}_{A}}{dt}\right)^{(2)}
S^{l}_{A}v^{j}_{A}\delta^{3}(\vec{x}-\vec{x}_{A})
+\sum_{A}\epsilon^{ijk}S^{k}_{A}
\left(\frac{\partial\phi_{A}}{\partial t}\right)
\delta^{3}(\vec{x}-\vec{x}_{A})
\nonumber\\
&&\ +\sum_{A}\epsilon^{(ikl}
\left(\frac{dS^{l}_{A}}{dt}\right)^{(2)}v^{k}_{A}v^{j)}_{A}
\delta^{3}(\vec{x}-\vec{x}_{A})
-\sum_{A}\epsilon^{ijk}
\left(\frac{dS^{k}_{A}}{dt}\right)^{(2)}\phi_{A}
\delta^{3}(\vec{x}-\vec{x}_{A})
\nonumber\\
&&\ +2\sum_{A}\epsilon^{(ikl}v^{j)}_{A}S^{l}_{A}\phi_{A}
\partial_{k}\delta^{3}(\vec{x}-\vec{x}_{A})
-\sum_{A}\epsilon^{(ikm}v^{k}_{A}S^{m}_{A}
v^{j)}_{A}v^{l}_{A}\partial_{l}\delta^{3}(\vec{x}-\vec{x}_{A})
\nonumber\\
&&\ +\sum_{A}\epsilon^{ijl}S^{l}_{A}v^{k}_{A}
(\partial^{A}_{k}\phi_{A})\delta^{3}(\vec{x}-\vec{x}_{A})
\nonumber\\
&&\ +2\sum_{A}\epsilon^{jkl}v^{k}_{A}S^{l}_{A}
(\partial^{A}_{i}\phi_{A})\delta^{3}(\vec{x}-\vec{x}_{A})
-2\sum_{A}\epsilon^{jkl}v^{i}_{A}S^{l}_{A}
(\partial^{A}_{k}\phi_{A})\delta^{3}(\vec{x}-\vec{x}_{A})
\nonumber\\
&&\ -\frac{1}{2}\sum_{A}\epsilon^{jkl}S^{l}_{A}
(\partial^{A}_{i}\zeta_{Ak})\delta^{3}(\vec{x}-\vec{x}_{A})
+\frac{1}{2}\sum_{A}\epsilon^{jkl}S^{l}_{A}
(\partial^{A}_{k}\zeta_{Ai})\delta^{3}(\vec{x}-\vec{x}_{A}),
\label{tij4}
\end{eqnarray}
where we have used the convention for symmetrization of indices,
\begin{equation}
A^{(i}B^{j)}\equiv\frac{1}{2}(A^{i}B^{j}+A^{j}B^{i})
\end{equation}

From this expansion of $T^{\mu\nu}$, the potentials $\phi$,
$\zeta_{i}$, and $\psi$ in Eqs.(\ref{phi}), (\ref{zeta}) 
and (\ref{psi}), respectively,
can be evaluated to be,
\begin{eqnarray}
\phi&=&-\sum_{A}\frac{m_{A}}{\vert\vec{x}-\vec{x}_{A}\vert},
\label{phip}
\\
\vec{\zeta}&=&-4\sum_{A}
\frac{m_{A}\vec{v}_{A}}{\vert\vec{x}-\vec{x}_{A}\vert}
+2\sum_{A}\frac{(\vec{x}-\vec{x}_{A})\times\vec{S}_{A}}
{\vert\vec{x}-\vec{x}_{A}\vert^{3}},
\label{zetap}
\\
\psi&=&\sum_{A}\sum_{B\neq A}
\frac{m_{A}}{\vert\vec{x}-\vec{x}_{A}\vert}
\frac{m_{B}}{\vert\vec{x}_{A}-\vec{x}_{B}\vert}
-\frac{1}{2}\sum_{A}\sum_{B\neq A}
\frac{m_{A}(\vec{x}-\vec{x}_{A})}{\vert\vec{x}-\vec{x}_{A}\vert}
\cdot
\frac{m_{B}(\vec{x}_{A}-\vec{x}_{B})}
{\vert\vec{x}_{A}-\vec{x}_{B}\vert^{3}}
\nonumber\\
&&\ +\frac{1}{2}\sum_{A}
\frac{m_{A}[\vec{v}_{A}\cdot(\vec{x}-\vec{x}_{A})]^{2}}
{\vert\vec{x}-\vec{x}_{A}\vert^{3}}
-2\sum_{A}\frac{m_{A}v^{2}_{A}}{\vert\vec{x}-\vec{x}_{A}\vert}
-2\sum_{A}\frac{(\vec{x}-\vec{x}_{A})\cdot
(\vec{v}_{A}\times\vec{S}_{A})}{\vert\vec{x}-\vec{x}_{A}\vert^{3}}.
\label{psip}
\end{eqnarray}

The equations of motion for $\vec{x}_{A}$ and $\vec{S}_{A}$ can
be derived, as discussed in Section II, by requiring $T^{\mu\nu}$
to be symmetric and conserving. For $T^{\mu\nu}$ to be symmetric,
we have the conditions,
\begin{eqnarray}
T^{0i}&=&T^{i0},
\label{sym1}
\\
T^{ij}&=&T^{ji}.
\label{sym2}
\end{eqnarray}
On the other hand, for $T^{\mu\nu}$ to be conserving, we have
\begin{eqnarray}
&&\nabla_{\mu}T^{\mu 0}=0
\nonumber\\
&\Rightarrow&\frac{\partial}{\partial t}T^{00}
+\partial_{i}T^{i0}+2\Gamma^{0}_{00}T^{00}+
3\Gamma^{0}_{0i}T^{i0}
+\Gamma^{i}_{i0}T^{00}+\Gamma^{0}_{ij}T^{ij}
+\Gamma^{i}_{ji}T^{i0}=0,
\label{con1}
\end{eqnarray}
and
\begin{eqnarray}
&&\nabla_{\mu}T^{\mu i}=0
\nonumber\\
&\Rightarrow&\frac{\partial}{\partial t}T^{0i}
+\partial_{j}T^{ij}+\Gamma^{0}_{00}T^{0i}+
\Gamma^{0}_{0j}T^{ij}+\Gamma^{k}_{k0}T^{0i}
\nonumber\\
&&\ \ \ \ \ +\Gamma^{k}_{kj}T^{ij}+\Gamma^{i}_{00}T^{00}
+2\Gamma^{i}_{0j}T^{0j}+\Gamma^{i}_{jk}T^{jk}=0.
\label{con2}
\end{eqnarray}
The lowest order of Eq.(\ref{sym1}) is satisfied automatically 
because $(T^{0i})^{(1)}=(T^{i0})^{(1)}$ as given in Eq.(\ref{t0i1}). 
From Eq.(\ref{sym2}), 
we have
\begin{eqnarray}
&&(T^{ij})^{(2)}=(T^{ji})^{(2)}
\nonumber\\
&\Rightarrow&
\left(\frac{d\vec{S}_{A}}{dt}\right)^{(2)}=0,
\label{eqs1}
\end{eqnarray}
which is the equation of motion for spins to the lowest order.
From Eq.(\ref{con1}), we have
\begin{equation}
\left(\frac{\partial}{\partial t}T^{00}\right)^{(0)}
+\partial_{i}(T^{0i})^{(1)}=0,
\end{equation}
which is also satisfied automatically, and so there is no new
constraint. From Eq.(\ref{con2}), 
\begin{eqnarray}
&&\left(\frac{\partial}{\partial t}T^{0i}\right)^{(1)}
+\partial_{j}(T^{ij})^{(2)}+(\Gamma^{i}_{00})^{(2)}(T^{00})^{(0)}=0
\nonumber\\
&\Rightarrow&\left(\frac{d\vec{v}_{A}}{dt}\right)^{(2)}
=-\vec{\nabla}^{A}\phi_{A}
=-\sum_{B\neq A}\frac{m_{B}\vec{r}_{AB}}{r_{AB}^{3}},
\label{eqp1}
\end{eqnarray}
where we have denoted $\vec{x}_{A}-\vec{x}_{B}$ by $\vec{r}_{AB}$.
This equation is just the Newtonian equation of motion for point
particles. The various components of the Christoffel symbol
$(\Gamma^{\mu}_{\alpha\beta})^{(i)}$ can be found in \cite{SW}. 

Now we go to the next order. 
From Eq.(\ref{sym1}), we have
\begin{equation}
(T^{0i})^{(3)}=(T^{i0})^{(3)},
\end{equation}
which is satisfied automatically as seen from Eqs.(\ref{t0i3})
and (\ref{eqs1}). 
Then from Eq.(\ref{sym2}), we have
\begin{eqnarray}
(T^{ij})^{(4)}&=&(T^{ji})^{(4)}
\nonumber\\
\Rightarrow\left(\frac{d\vec{S}^{A}}{dt}\right)^{(4)}
&=&\left(2\left(\frac{\partial\phi_{A}}{\partial t}\right)
+\vec{v}_{A}\cdot\vec{\nabla}_{A}\phi_{A}\right)\vec{S}_{A}
-(\vec{S}_{A}\cdot\vec{v}_{A})\vec{\nabla}_{A}\phi_{A}
\nonumber\\
&&\ +2(\vec{S}_{A}\cdot\vec{\nabla}_{A}\phi_{A})\vec{v}_{A}
-\frac{1}{2}(\vec{S}_{A}\cdot\vec{\nabla}_{A})\vec{\zeta}_{A}
+\frac{1}{2}\vec{\nabla}_{A}(\vec{S}_{A}\cdot\vec{\zeta}_{A})
\nonumber\\
&=&\sum_{B\neq A}\frac{m_{B}\vec{r}_{AB}\cdot
(\vec{v}_{A}-2\vec{v}_{B})}{r^{3}_{AB}}\vec{S}_{A}
-\sum_{B\neq A}
\frac{m_{B}[(\vec{v}_{A}-2\vec{v}_{B})\cdot\vec{S}_{A}]}
{r^{3}_{AB}}\vec{r}_{AB}
\nonumber\\
&&\ +2\sum_{B\neq A}\frac{m_{B}(\vec{r}_{AB}\cdot\vec{S}_{A})}
{r^{3}_{AB}}(\vec{v}_{A}-\vec{v}_{B})
-2\sum_{B\neq A}\frac{\vec{S}_{A}\times\vec{S}_{B}}
{r^{3}_{AB}}
\nonumber\\
&&\ +3\sum_{B\neq A}\frac{\vec{r}_{AB}\cdot
(\vec{S}_{A}\times\vec{S}_{B})}{r^{5}_{AB}}\vec{r}_{AB}
+3\sum_{B\neq A}\frac{(\vec{r}_{AB}\cdot\vec{S}_{A})}
{r^{5}_{AB}}(\vec{r}_{AB}\times\vec{S}_{B}),
\label{eqs2}
\end{eqnarray}
where we have used the results in Eqs.(\ref{phip}) to 
(\ref{psip}) for the potentials.

Next from Eq.(\ref{con1}), 
\begin{eqnarray}
&&\left(\frac{\partial}{\partial t}T^{00}\right)^{(2)}
+\partial_{i}(T^{0i})^{(3)}+2(\Gamma^{0}_{00})^{(3)}(T^{00})^{(0)}
\nonumber\\
&&\ \ +3(\Gamma^{0}_{0i})^{(2)}(T^{i0})^{(1)}
+(\Gamma^{i}_{i0})^{(3)}(T^{00})^{(0)}
+(\Gamma^{j}_{ji})^{(2)}(T^{0i})^{(1)}=0,
\end{eqnarray}
which is satisfied automatically because of the Newtonian equation
of motion as in Eq.(\ref{eqp1}). Then from Eq.(\ref{con2}),
\begin{eqnarray}
&&\left(\frac{\partial}{\partial t}T^{0i}\right)^{(4)}
+\partial_{j}(T^{ij})^{(4)}+(\Gamma^{0}_{00})^{(3)}(T^{0i})^{(1)}
+(\Gamma^{0}_{0j})^{(2)}(T^{ij})^{(2)}
\nonumber\\
&&\ \ +(\Gamma^{k}_{k0})^{(3)}(T^{0i})^{(1)}
+(\Gamma^{k}_{kj})^{(2)}(T^{ij})^{(2)}
+(\Gamma^{i}_{00})^{(2)}(T^{00})^{(2)}
\nonumber\\
&&\ \ \ \ +(\Gamma^{i}_{00})^{(4)}(T^{00})^{(0)}
+2(\Gamma^{i}_{0j})^{(3)}(T^{0j})^{(1)}
+(\Gamma^{i}_{jk})^{(2)}(T^{jk})^{(2)}=0,
\end{eqnarray}
which gives
\begin{eqnarray}
\left(\frac{d\vec{v}_{A}}{dt}\right)^{(4)}&=&
3\vec{v}_{A}\left(\frac{\partial\phi_{A}}{\partial t}\right)
-v^{2}_{A}(\vec{\nabla}_{A}\phi_{A})
+4\vec{v}_{A}(\vec{v}_{A}\cdot\vec{\nabla}_{A}\phi_{A})
\nonumber\\
&&\ -4\phi_{A}(\vec{\nabla}_{A}\phi_{A})
-\vec{\nabla}_{A}\psi_{A}
-\frac{\partial\vec{\zeta}_{A}}{\partial t}
+\vec{v}_{A}\times(\vec{\nabla}_{A}\times\vec{\zeta}_{A})
\nonumber\\
&&\ +\frac{1}{m_{A}}\vec{S}_{A}\times
\left(\vec{\nabla}_{A}
\left(\frac{\partial\phi_{A}}{\partial t}\right)\right)
+\frac{1}{m_{A}}\vec{S}_{A}\times\vec{\nabla}_{A}
(\vec{v}_{A}\cdot\vec{\nabla}_{A}\phi_{A})
\nonumber\\
&&\ -\frac{2}{m^{A}}\vec{\nabla}_{A}
\left[(\vec{v}_{A}\times\vec{S}_{A})
\cdot\vec{\nabla}_{A}\phi_{A}\right]
+\frac{1}{2m_{A}}\vec{\nabla}\left[\vec{S}_{A}\cdot
(\vec{\nabla}_{A}\times\vec{\zeta}_{A})\right]
\nonumber\\
&=&\sum_{B\neq A}\frac{m_{B}\vec{r}_{AB}}{r^{3}_{AB}}
\left[4\sum_{C\neq A}\frac{m_{C}}{r_{AC}}
+\sum_{C\neq A,B}\frac{m_{C}}{r_{BC}}
\left(1-\frac{\vec{r}_{AB}\cdot\vec{r}_{BC}}{r^{2}_{BC}}\right)
\right.
\nonumber\\
&&\ \ \ \ \ \left.+5\frac{m_{A}}{r_{AB}}-v^{2}_{A}
+4\vec{v}_{A}\cdot\vec{v}_{B}-2v^{2}_{B}
+\frac{3}{2}\frac{(\vec{r}_{AB}\cdot\vec{v}_{B})^{2}}
{r^{2}_{AB}}\right]
\nonumber\\
&&\ -\frac{7}{2}\sum_{B\neq A}\sum_{C\neq A,B}
\frac{m_{C}\vec{r}_{BC}}{r^{3}_{BC}}\frac{m_{B}}{r_{AB}}
+\sum_{B\neq A}\frac{m_{B}\vec{r}_{AB}\cdot
(4\vec{v}_{A}-3\vec{v}_{B})}{r^{3}_{AB}}(\vec{v}_{A}-\vec{v}_{B})
\nonumber\\
&&\ +6\sum_{B\neq A}\frac{\vec{r}_{AB}}{m_{A}r^{3}_{AB}}
\left[\frac{[\vec{r}_{AB}\times(\vec{v}_{A}-\vec{v}_{B})]\cdot
(m_{A}\vec{S}_{B}+m_{B}\vec{S}_{A})}{r^{2}_{AB}}
-\frac{\vec{S}_{A}\cdot\vec{S}_{B}}{r^{2}_{AB}}
\right.
\nonumber\\
&&\left.\ \ \ \ \ 
+\frac{5}{2}\frac{(\vec{r}_{AB}\cdot\vec{S}_{A})
(\vec{r}_{AB}\cdot\vec{S}_{B})}{r^{4}_{AB}}\right]
\nonumber\\
&&\ -3\sum_{B\neq A}\frac{1}{m_{A}r^{5}_{AB}}
\left[(\vec{r}_{AB}\cdot\vec{S}_{A})\vec{S}_{B}
+(\vec{r}_{AB}\cdot\vec{S}_{B})\vec{S}_{A}\right]
\nonumber\\
&&\ -\sum_{B\neq A}\frac{1}{m_{A}r^{3}_{AB}}(\vec{v}_{A}-\vec{v}_{B})
\times(4m_{A}\vec{S}_{B}+3m_{B}\vec{S}_{A})
\nonumber\\
&&\ +3\sum_{B\neq A}\frac{1}{m_{A}r^{5}_{AB}}
\vec{r}_{AB}\cdot(\vec{v}_{A}-\vec{v}_{B})
\left[\vec{r}_{AB}\times(2m_{A}\vec{S}_{B}+m_{B}\vec{S}_{A})\right].
\label{eqp2}
\end{eqnarray}
Hence, Eqs.(\ref{eqs1}), (\ref{eqp1}), (\ref{eqs2}), and (\ref{eqp2}) 
form a complete set of equations of motion
for $\vec{x}_{A}(t)$ and $\vec{S}_{A}(t)$ for spinning particles to the 
(post)$^{2}$-Newtonian order. This can be considered as an extension to the
well-known Einstein-Infeld-Hoffmann equations \cite{EIH} for ordinary 
particles. 

Using these equations of motion, the expressions for the energy-momentum
tensor $T^{\mu\nu}$ in Eqs.(\ref{t000}) to (\ref{tij4}) 
can be simplified to
\begin{eqnarray}
(T^{00})^{(0)}&=&\sum_{A}m_{A}\delta^{3}(\vec{x}-\vec{x}_{A}),
\\
(T^{00})^{(2)}&=&\sum_{A}m_{A}(\phi_{A}+\frac{1}{2}v^{2}_{A})
\delta^{3}(\vec{x}-\vec{x}_{A})
-\sum_{A}\epsilon^{ijk}v^{j}_{A}S^{k}_{A}\partial_{i}
\delta^{3}(\vec{x}-\vec{x}_{A}),
\\
(T^{0i})^{(1)}&=&(T^{i0})^{(1)}
\nonumber\\
&=&\sum_{A}m_{A}v^{i}_{A}\delta^{3}(\vec{x}-\vec{x}_{A})
+\frac{1}{2}\sum_{A}\epsilon^{ijk}S^{k}_{A}\partial_{j}
\delta^{3}(\vec{x}-\vec{x}_{A})
\\
(T^{0i})^{(3)}&=&(T^{i0})^{(3)}
\nonumber\\
&=&\sum_{A}m_{A}(\phi_{A}+\frac{1}{2}v^{2}_{A})v_{A}^{i}
\delta^{3}(\vec{x}-\vec{x}_{A})
\nonumber\\
&&\ +\sum_{A}\epsilon^{ijk}S^{j}_{A}(\partial^{A}_{k}\phi_{A})
\delta^{3}(\vec{x}-\vec{x}_{A})
-\sum_{A}\epsilon^{ijk}S^{j}_{A}\phi_{A}\partial_{k}
\delta^{3}(\vec{x}-\vec{x}_{A})
\nonumber\\
&&\ -\frac{1}{2}\sum_{A}\epsilon^{ikl}v^{j}_{A}v^{k}_{A}S^{l}_{A}
\partial_{j}\delta^{3}(\vec{x}-\vec{x}_{A})
-\frac{1}{2}\sum_{A}\epsilon^{jkl}v^{i}_{A}v^{k}_{A}S^{l}_{A}
\partial_{j}\delta^{3}(\vec{x}-\vec{x}_{A}),
\\
(T^{ij})^{(2)}&=&\sum_{A}m_{A}v^{i}_{A}v^{j}_{A}
\delta^{3}(\vec{x}-\vec{x}_{A})
+\sum_{A}\epsilon^{(ikl}v^{j)}_{A}S^{l}_{A}\partial_{k}
\delta^{3}(\vec{x}-\vec{x}_{A}),
\\
(T^{ij})^{(4)}&=&\sum_{A}m_{A}(\phi_{A}+\frac{1}{2}v^{2}_{A})
v^{i}_{A}v^{j}_{A}\delta^{3}(\vec{x}-\vec{x}_{A})
\nonumber\\
&&\ +2\sum_{A}\epsilon^{(ikl}v^{j)}_{A}S^{l}_{A}\phi_{A}\partial_{k}
\delta^{3}(\vec{x}-\vec{x}_{A})
-4\sum_{A}\epsilon^{(ikl}v^{j)}_{A}S^{l}_{A}(\partial^{A}_{k}
\phi_{A})\delta^{3}(\vec{x}-\vec{x}_{A})
\nonumber\\
&&\ +\sum_{A}\epsilon^{(ikl}v^{k}_{A}S^{l}_{A}(\partial^{j)}_{A}
\phi_{A})\delta^{3}(\vec{x}-\vec{x}_{A})
-\sum_{A}\epsilon^{(ikm}v^{j)}_{A}v^{k}_{A}S^{m}_{A}
v^{l}_{A}\partial_{l}\delta^{3}(\vec{x}-\vec{x}_{A})
\nonumber\\
&&\ -\frac{1}{2}\sum_{A}\epsilon^{(ikl}S^{l}_{A}(\partial^{j)}_{A}
\zeta_{Ak})\delta^{3}(\vec{x}-\vec{x}_{A})
+\frac{1}{2}\sum_{A}\epsilon^{(ikl}S^{l}_{A}(\partial^{A}_{k}
\zeta^{j)}_{A})\delta^{3}(\vec{x}-\vec{x}_{A}),
\end{eqnarray}
and the expressions for the potentials $\phi_{A}$ and $\zeta^{i}_{A}$
are those given in Eqs.(\ref{phip}) and (\ref{zetap}). 
These components of the energy-momentum
tensor $T^{\mu\nu}$ to various post-Newtonian order will be used in the
next section to find the gravitational wave generated by a
system of N spinning particles in the post-Newtonian approximation.

\section{Gravitational waves generation}

In this section we follow the works of EW \cite{EW} and
Wagoner and Will \cite{WW} to consider the gravitational wave generation
of N spinning particles in the post-Newtonian approximation. 
As in \cite{EW}, we derive the 
waveform and the luminosity of the gravitational waves up to 
post-Newtonian order.

The waveform of the gravitational wave in the radiation zone
is given by \cite{EW}
\begin{equation}
h^{ij}_{TT}=\frac{2}{R}\frac{\partial^{2}}{\partial t^{2}}
\left[I^{ij}(t-R)+n_{k}I^{ijk}(t-R)+n_{k}n_{l}I^{ijkl}(t-R)
\right]_{TT},
\end{equation}
where $TT$ denotes the transverse-traceless part, 
$n_{k}\equiv R_{k}/R$, and
\begin{eqnarray}
I^{ij}&=&\int\tau^{00}x^{i}x^{j}d^{3}\!x,
\label{ij}
\\
I^{ijk}&=&\int (2\tau^{0(i}x^{j)}x^{k}-\tau^{0k}x^{i}x^{j})d^{3}\!x,
\label{ijk}
\\
I^{ijkl}&=&\int\tau^{ij}x^{k}x^{l}d^{3}\!x,
\label{ijkl}
\end{eqnarray}
where $\tau^{\mu\nu}$ is the total energy-momentum tensor,
\begin{equation}
\tau^{\mu\nu}=T^{\mu\nu}+t^{\mu\nu},
\end{equation}
which includes the matter part $T^{\mu\nu}$ and the gravitational
part $t^{\mu\nu}$. 

Here, for spinning particles, we use the energy-momentum tensor
$T^{\mu\nu}$
discussed in the last section. While $t^{\mu\nu}$, up to post-Newtonian
order, is given in \cite{WW}. Therefore, 
\begin{eqnarray}
\tau^{00}&=&\sum_{A}m_{A}(1+\phi_{A}+\frac{1}{2}v^{2}_{A})
\delta^{3}(\vec{x}-\vec{x}_{A})
-\sum_{A}\epsilon^{ijk}v^{j}_{A}S^{k}_{A}\partial_{i}
\delta^{3}(\vec{x}-\vec{x}_{A})
\nonumber\\
&&\ -\frac{1}{8\pi}(4\phi\partial^{2}\phi+
3\partial_{i}\phi\partial_{i}\phi),
\label{tau00}
\\
\tau^{0i}&=&\sum_{A}m_{A}v^{i}_{A}\delta^{3}(\vec{x}-\vec{x}_{A})
-\frac{1}{2}\sum_{A}\epsilon^{ijk}S^{j}_{A}\partial_{k}
\delta^{3}(\vec{x}-\vec{x}_{A}),
\label{tau0i}
\\
\tau^{ij}&=&\sum_{A}m_{A}v^{i}_{A}v^{j}_{A}
\delta^{3}(\vec{x}-\vec{x}_{A})
+\sum_{A}\epsilon^{(ikl}S^{l}_{A}v^{j)}_{A}\partial_{k}
\delta^{3}(\vec{x}-\vec{x}_{A})
\nonumber\\
&&\ +\frac{1}{8\pi}\left[-2\partial_{i}\phi
\partial_{j}\phi-4\phi\partial_{i}\partial_{j}\phi
+\delta_{ij}(4\phi\partial^{2}\phi+3\partial_{k}\phi
\partial_{k}\phi)\right],
\label{tauij}
\end{eqnarray}
up to post-Newtonian order,
where the last terms in Eq.(\ref{tau00}) and Eq.(\ref{tauij}) 
come from the contributions of 
$t^{00}$ and $t^{ij}$, respectively. Then the integrals in 
Eqs.(\ref{ij}) to (\ref{ijkl}) can be evaluated,
\begin{eqnarray}
I^{ij}&=&\sum_{A}m_{A}x^{i}_{A}x^{j}_{A}\left(
1+\frac{1}{2}v^{2}_{A}-\frac{1}{2}\sum_{B\neq A}
\frac{m_{B}}{r_{AB}}\right)
+2\sum_{A}\epsilon^{(ikl}v^{k}_{A}S^{l}_{A}x^{j)},
\label{ijp}
\\
I^{ijk}&=&\sum_{A}m_{A}(2v^{(i}_{A}x^{j)}_{A}x^{k}_{A}
-v^{k}_{A}x^{i}_{A}x^{j}_{A})
+2\sum_{A}\epsilon^{k(il}x^{j)}_{A}S^{l}_{A},
\label{ijkp}
\\
I^{ijkl}&=&\sum_{A}m_{A}v^{i}_{A}v^{j}_{A}x^{k}_{A}x^{l}_{A}
-\frac{1}{12}\sum_{A}\sum_{B\neq A}
\frac{m_{A}m_{B}r^{i}_{AB}r^{j}_{AB}}{r_{AB}}
\left[\delta^{kl}+\frac{1}{r_{AB}}(-r^{k}_{AB}r^{l}_{AB}
+6x^{k}_{A}x^{l}_{A})\right]
\nonumber\\
&&\ -\sum_{A}\left(\epsilon^{(ikm}v^{j)}_{A}x^{l}_{A}
+\epsilon^{(ilm}v^{j)}_{A}x^{k}_{A}\right)S^{m}_{A}.
\label{ijklp}
\end{eqnarray}
In obtaining these expressions, we have discarded terms with vanishing
transverse-traceless parts, of which some are actually divergent,
because those terms will not contribute
to the generation of gravitational waves \cite{WW}. For example,
\begin{eqnarray}
\left(\delta_{ij}\right)_{TT}=0,
\\
\left(n^{i}f^{j}\right)_{TT}=0.
\end{eqnarray}
for some function $f^{j}$. Also we have used the
formula,
\begin{eqnarray}
&&\int\phi\partial_{i}\partial_{j}\phi x^{k}x^{l}d^{3}\!x
\nonumber\\
&=&\sum_{A}\sum_{B\neq A}\frac{\pi m_{A}m_{B}r^{i}_{AB}r^{j}_{AB}}
{3r^{AB}}\left[\delta^{kl}+\frac{1}{r^{2}_{AB}}(2r^{k}_{AB}r^{l}_{AB}
-6r^{(k}_{AB}x^{l)}_{A}+6x^{k}_{A}x^{l}_{A})\right],
\end{eqnarray}
which can be proved \cite{WW} by ignoring terms which have no
transverse-traceless parts in a similar way.

Putting these together the waveform of the gravitational wave,
up to post-Newtonian order, is
\begin{eqnarray}
h^{ij}_{TT}=\frac{2}{R}\frac{\partial^{2}}{\partial t^{2}}
&&\left\{\sum_{A}m_{A}x^{i}_{A}x^{j}_{A}
\left(1-\hat{n}\cdot\vec{v}_{A}+\frac{1}{2}v^{2}_{A}
-\frac{1}{2}\sum_{B\neq A}\frac{m^{B}}{r^{AB}}\right)
\right.
\nonumber\\
&&\ +2\sum_{A}m_{A}v^{(i}_{A}x^{j)}_{A}(\hat{n}\cdot\vec{x}_{A})
+\sum_{A}m_{A}v^{i}_{A}v^{j}_{A}(\hat{n}\cdot\vec{x}_{A})^{2}
\nonumber\\
&&\ -\frac{1}{12}\sum_{A}\sum_{B\neq A}
\frac{m_{A}m_{B}r^{i}_{AB}r^{j}_{AB}}{r_{AB}}
\left[1+\frac{1}{r^{2}_{AB}}\left(-(\hat{n}\cdot\vec{r}_{AB})^{2}
+6(\hat{n}\cdot\vec{x}_{A})^{2}\right)\right]
\nonumber\\
&&\left.\ +2\sum_{A}x^{(i}_{A}(\vec{v}_{A}\times\vec{S}_{A})^{j)}
-2\sum_{A}x^{(i}_{A}(\hat{n}\times\vec{S}_{A})^{j)}
-2\sum_{A}(\hat{n}\cdot\vec{x}_{A})v^{(i}_{A}
(\hat{n}\times\vec{S}_{A})^{j)}\right\}_{TT}.
\nonumber\\
&&
\end{eqnarray}
Hence, together with the equations of motion in 
Eqs.(\ref{eqs1}), (\ref{eqp1}), 
(\ref{eqs2}) and (\ref{eqp2}), one can
obtain the gravitational waves generated by N spinning particles.

The total luminosity of the gravitational wave is also given in
\cite{EW} as,
\begin{eqnarray}
L&=&\langle\frac{1}{5}{\cal N}_{ij}{\cal N}_{ij}
+\frac{1}{105}(11{\cal N}_{ijk}{\cal N}_{ijk}
-6{\cal N}_{ijj}{\cal N}_{ikk}
\nonumber\\
&&\ \ \ \ \ 
-6{\cal N}_{ijk}{\cal N}_{ikj}+22{\cal N}_{ij}{\cal N}_{ijkk}
-24{\cal N}_{ij}{\cal N}_{ikkj})+\cdots\rangle,
\end{eqnarray}
where the bracket denotes an average over several characteristic
wavelengths, and
\begin{equation}
{\cal N}_{ijk_{1}\cdots k_{m}}\equiv
\frac{d^{3}}{dt^{3}}\left(I_{ijk_{1}\cdots k_{m}}
-\frac{1}{3}\delta^{ij}I_{llk_{1}\cdots k_{m}}\right).
\end{equation}
Up to post-Newtonian order, $L$ can be expanded to
\begin{eqnarray}
L&=&\frac{1}{5}\langle{\cal N}_{ij}^{(0)}{\cal N}_{ij}^{(0)}
+\frac{1}{21}(42{\cal N}_{ij}^{(0)}{\cal N}_{ij}^{(2)}
+11{\cal N}_{ijk}^{(1)}{\cal N}_{ijk}^{(1)}
-6{\cal N}_{ijj}^{(1)}{\cal N}_{ikk}^{(1)}
\nonumber\\
&&\ -6{\cal N}_{ijk}^{(1)}{\cal N}_{ikj}^{(1)}
+22{\cal N}_{ij}^{(0)}{\cal N}_{ijkk}^{(2)}
-24{\cal N}_{ij}^{(0)}{\cal N}_{ikkj}^{(2)})\rangle.
\end{eqnarray}
Using the results for $I^{ij}$, $I^{ijk}$, and $I^{ijkl}$ in 
Eqs.(\ref{ijp}), (\ref{ijkp}), and (\ref{ijklp}), respectively, 
one can obtain the various orders of 
${\cal N}_{ijk_{1}\cdots k_{m}}$ as follows.
\begin{eqnarray}
{\cal N}_{ij}^{(0)}&=&\frac{d^{3}}{dt^{3}}
\left\{\sum_{A}m_{A}\left(x^{i}_{A}x^{j}_{A}
-\frac{1}{3}\delta^{ij}x^{2}_{A}\right)\right\},
\\
{\cal N}_{ij}^{(2)}&=&\frac{d^{3}}{dt^{3}}
\left\{\sum_{A}m_{A}\left(x^{i}_{A}x^{j}_{A}-
\frac{1}{3}\delta^{ij}x^{2}_{A}\right)
\left(\frac{1}{2}v^{2}_{A}
-\frac{1}{2}\sum_{B\neq A}\frac{m^{B}}{r_{AB}}\right)
\right.\nonumber\\
&&\left.\ +2\sum_{A}\left[x^{(i}_{A}(\vec{v}_{A}\times\vec{S}_{A})^{j)}
-\frac{1}{3}\delta^{ij}\vec{x}_{A}\cdot
(\vec{v}_{A}\times\vec{S}_{A})\right]\right\},
\\
{\cal N}_{ijk}^{(1)}&=&\frac{d^{3}}{dt^{3}}
\left\{2\sum_{A}m_{A}\left(v^{(i}_{A}x^{j)}_{A}
-\frac{1}{3}\delta^{ij}\vec{v}_{A}\cdot\vec{x}_{A}\right)x^{k}_{A}
-\sum_{A}m_{A}\left(x^{i}_{A}x^{j}_{A}
-\frac{1}{3}\delta^{ij}x^{2}_{A}\right)v^{k}_{A}
\right.\nonumber\\
&&\left.\ +2\sum_{A}\left[\epsilon^{k(il}x^{j)}_{A}S^{l}_{A}
-\frac{1}{3}\delta^{ij}(\vec{x}_{A}\times\vec{S}_{A})^{k}\right]
\right\},
\\
{\cal N}_{ijkl}^{(2)}&=&\frac{d^{3}}{dt^{3}}\left\{
\sum_{A}m_{A}\left(v^{i}_{A}v^{j}_{A}
-\frac{1}{3}\delta^{ij}v^{2}_{A}\right)x^{k}_{A}x^{l}_{A}
\right.\nonumber\\
&&\ -\frac{1}{12}\sum_{A}\sum_{B\neq A}\frac{m_{A}m_{B}}{r_{AB}}
\left(r^{i}_{AB}r^{j}_{AB}-\frac{1}{3}\delta^{ij}r^{2}_{AB}\right)
\left[\delta^{kl}+\frac{1}{r^{2}_{AB}}
\left(-r^{k}_{AB}r^{l}_{AB}+6x^{k}_{A}x^{l}_{A}\right)\right]
\nonumber\\
&&\left.\ \sum_{A}\left[\epsilon^{l(im}v^{j)}_{A}S^{m}_{A}
-\frac{1}{3}\delta^{ij}(\vec{v}_{A}\times\vec{S}_{A})^{l}\right]
x^{k}_{A}+\sum_{A}\left[\epsilon^{k(im}v^{j)}_{A}S^{m}_{A}
-\frac{1}{3}\delta^{ij}(\vec{v}_{A}\times\vec{S}_{A})^{k}\right]
x^{l}_{A}\right\}.
\nonumber\\
&&
\end{eqnarray}
This completes our consideration of the gravitational wave 
generated by the motion of N spinning particles to the
post-Newtonian order.

\section{Conclusions and discussions}

In this work we consider a method to develop the post-Newtonian
expansion for both the equations of motion and the generation of
gravitational wave of N spinning particles. In this approach the
energy-momentum tensor $T^{\mu\nu}$
for spinning particles, due to Dixon and
BI, is introduced first. Using this $T^{\mu\nu}$, 
one can derive the Papapetrou equations by requiring
it to be symmetric and conserving. By the formalism of
EW, we can then obtain the waveform and the luminosity of the
gravitational wave generated by the motion of these spinning particles.

This approach is straight forward and much simpler than the BDI
multipole formalism since we still regard the particles as point-like.
It is self-contained in the sense that once we write down $T^{\mu\nu}$,
the equations of motion can also be derived from it. Another merit 
is that the BI formalism is very general so that structures 
other than spin, such as tidal effects, can also be taken
into account systematically by introducing terms corresponding to these
structures in $T^{\mu\nu}$. 

However, there are several points in our calculation that we must 
be careful with. First, the EW formalism is less rigorous
than the BDI formalism in which integrals like those in 
Eqs.(\ref{ij}) to (\ref{ijkl}) 
have divergent parts. Fortunately, these divergent parts have no
transverse-traceless parts so they will not 
contribute to the generation of gravitational waves. We therefore expect
the two formalisms to give the same results.

Another point to note is that the harmonic coordinate condition
used in Section III, as well as in \cite{SW}, is not the same as the
gauge condition used in the EW formalism. It is nevertheless
shown in \cite{AGW} that, at least to the post-Newtonian order that 
we are considering here, the extra terms coming from these different
gauge choices all have vanishing transverse-traceless parts. Therefore,
there is no inconsistency in our investigation in this paper concerning 
gauge conditions. On the other hand, 
if we want to extend our consideration to 
higher post-Newtonian orders, we must be more careful with gauge choices.

In our calculation we have assumed that spin $\vec{S}$ is of post-Newtonian
order, same as velocity $\vec{v}$. Then the spin-orbit and spin-spin
effects come in both at the post-Newtonian order. Since these effects
for compact objects come in at even higher orders, we are really dealing
with ordinary objects. We have chosen this simpler case just as an
illustration of how our formalism works for spinning particles while
staying at the post-Newtonian level. Of course, the compact case is 
more interesting because of the possibility of detection
of their coalescences by LIGO and VIRGO in the near future. We plan
to consider that situation in a separate work.

\acknowledgements

This work is supported by the National Science Council of the
Republic of China under contract number NSC 85-2112-M032-001.

\end{document}